# Effect of ELF e.m. fields on metalloprotein redox –active sites

De Ninno A.[1], Prosdocimi M[2]., Ferrari V.[3], Gerardi G.[3], Barbaro F.[2], Badon T.[3], Bernardini D.[3]

(1) ENEA, CR Frascati, Dept. FIM,[1] Frascati (Rome), Italy
(2) PROMETEO S.r.l., Padua, Italy
(3) Department of Veterinary Clinical Sciences,–University of Padua, Italy

**Abstract:** The peculiarity of the distribution and geometry of metallic ions in enzymes pushed us to set the hypothesis that metallic ions in active-site act like tiny antennas able to pick up very feeble e.m. signals. Enzymatic activity of $Cu^{2+}$, $Zn^{2+}$ Superoxide Dismutase (SOD1) and $Fe^{2+}$ Xanthine Oxidase (XO) has been studied, following in vitro generation and removal of free radicals. We observed that Superoxide radicals generation by XO is increased by a weak field having the Larmor frequency $f_L$ of $Fe^{2+}$ while the SOD1 kinetics is sensibly reduced by exposure to a weak field having the frequency $f_L$ of $Cu^{2+}$ ion.

Abbreviations:
SOD Superoxide Dismutase
SOD1 Cu-Zn Superoxide Dismutase
XO Xanthine Oxidase
ELF Extremely Low frequency

---

[1]*Corresponding Author: Via E. Fermi 45 00044 Frascati (Rome) - Italy.Tel: +39 0694005792, Fax: +39 0694005011 e-mail: deninno@frascati.enea.it*

# 1. Introduction

The detailed mechanism of enzymatic catalysis taking place in living organism is still hardly understood. Though many details in the protein structure are well known, it is very difficult to build a picture of static and even more of dynamic information about protein conformation (folding) and activity, and thus despite the large experimental and theoretical efforts in this field. The protein conformation was investigated especially through very sophisticated Monte Carlo simulation and molecular dynamics. However, these successful models are merely descriptive since they are not able to justify neither the conformation nor the enzymatic activity of the proteins [1].

In the 80s it was recognised [2, 3] that very weak alternating magnetic fields applied to living organisms produce variations in the ion concentrations within the cells when the frequency of the applied field matches the angular frequency $f_L = \frac{1}{2\pi} e/m \, B_0$, being $m$ the mass, $e$ the charge of the ion and $B_0$ the amplitude of a static additional magnetic field. The typical frequency range, for most of the ions involved, lies between 10 and 50 Hz.

The simplest coupling model toy of the motion of a single particle of mass $m$ and charge $e$ in a constant magnetic field of strength $B_0$ leads to the superposition of a uniform precession of angular frequency $f_L$ (known as Larmor precession) about the direction of the field on the original motion. The Larmor precession provides a mechanism by which biological systems become sensitive to small static and resonating magnetic fields [4]. However, it just faces the first step of the very complex problem of the enzymatic catalysis: how the signal picked up by the ions could propagate the information to the whole structure of the enzyme remains, at present, still unknown. Moreover it has been pointed out by Adair [5,6] that in a gas-like model the average rate of collisions among particles and the energy transfer at room temperature prevents the classical angular precession and destroys any information carried by the field. However, as clearly stated by Edmonds [4], the peculiarity of Larmor precession makes the existence of a resonating effect also at room temperature possible, provided that ions are considered as tightly bounded in a central force field, as they actually are, and not so free as particles in a gas-like model. Further, the perturbation introduced by the e.m. field must be negligible. This sets an upper limit to the intensity of the energy transferred to the ions by the external field and hence to the amplitude of the e.m. perturbation. An additional problem, to be extensively investigated, is how the ion precession can affect the active site geometry and then influence the whole enzyme dynamics.

Also the interaction of spin-correlated radical pairs with magnetic fields is a possible magnetosensitive effect [7,8,9] to account for the above phenomena. However, contrary to the Larmor frequencies, the range of effective frequencies for organic radicals lies in the 1-100 Mhz range.

Our attention was focused on Superoxide Dismutases (SOD) enzymes as model system in order to investigate the effect of the coupling of a resonating e.m. field to a catalytic reaction "in vitro", and to check whether the enzyme active site couples with the magnetic field according to Edmonds' proposal.

SODs are metalloenzymes that catalyze the dismutation of superoxide $O_2^-$ to molecular oxygen forming $H_2O_2$ and then, with the help of catalase, produce $H_2O$ [10]. They are widely distributed in animal and vegetal world. Superoxide is generated in a great number of cellular processes and in respiration. The structure of Cu-Zn SOD1 is well known [11]: it is a stable homodimer with two subunit. It contains 2632 atoms and its weight is 15857.8 a.m.u. Each monomer just contains two metal ions, one copper and one zinc, weight respectively 63.55 and 65.38 a.m.u., which play a crucial structural and catalytic role in spite of their scarcity. The Cu is bound by four histidines in two different geometries (tetrahedral and planar trigonal), according to its oxidation state.

The conventional description of the reaction is the following:

$$M^{2+} + O_2^- \rightarrow M^+ + O_2$$

$$M^+ + O_2^- + 2H^+ \rightarrow M^{2+} + H_2O_2$$

The mechanism of the SOD1 activity has been widely described by R. Rakhit and A. Chakrabartty, [12] and a tentative hypothesis on its dynamics has been provided by P.J. Hart et al., [13]. The first step of the reaction is the binding of oxygen to the active site: here the electron transfer to Cu occurs and then oxygen diffuses out. This causes the spatial rearrangement of Cu to the trigonal state which is then oxidized by a second, not bounded superoxide (the electron is transferred through space), thus regenerating the initial tetrahedral Cu state. The Edmonds' proposal is relevant to a single particle only subject to a centrally directed force, when a constant magnetic field is applied on Z axis. The actual geometry of the Cu active site is rather more complicate. Ricardo J. and coworkers [14] showed that the copper ion has no first solvation shell and that the second solvation shell is placed at about 4-7 Å from the copper ion, thus generating an empty cavity. Such a cavity is not spherical as it can be seen by the calculated metal-ligand bonding parameters. Hence it turns out that the Edmond model should be completed including higher order effects, which are out of the scope of this paper. Such an observation should lead to envisaging an interaction more complex than the simple resonating Larmor effect.

At present time the hypothesis can only be formulated that a physical mechanism of enzyme activation exists based on e.m. coupling of the charged metallic ions and a suitable field. Active-sites can act like tiny antennas able to pick up very feeble e.m. signals.

## 2. Materials and Methods

The effect of an exposure to electromagnetic fields was measured in two different setup, both characterised by measurement of the absorbance of a solution containing a tetrazolium salt. Superoxide radicals generation (caused by hypoxanthine oxidation catalysed by Xanthine Oxidase) is proportional to the reduction of tetrazolium salt to formazan, directly estimated by measuring the increased absorbance values at 450 nm. For our measurements we always utilised reagents commercialised by Cayman Chemical as a Superoxide Dismutase Assay Kit (supplied by Prodotti Gianni, Milano, Italy), performing the assays measuring absorbance at room temperature at different time intervals. Assay mixture contained a final volume of 230 micro litres, prepared by mixing 200 micro litres of assay buffer (containing diethylenetriaminepentaacetic acid, the tetrazolium salt and hypoxanthine) with 10 micro litres of a solution of different content and 20 micro litres of a solution of XO, added to start the reaction.

The content of the different assay mixtures was as follows: 1) determination of XO activity: total reaction volume was completed with 10 micro litres of distilled water; 2) determination of XO/SOD activity: total reaction volume was completed with 10 micro litres of a SOD solution having a final activity of 0.15 U/ml. SOD (Cu/Zn) was a bovine erythrocyte preparation.

Each assay was incubated in three wells in a plate and processed in a plate reader TECAN mod. Sunrise equipped with software MAGELLAN V 3.11 (TECAN ITALIA). Each well was read independently so that each point on the kinetic curve is the average of three independent lectures. We run control test before and after each experimental session by using the same assay used for the exposure. The magnetic field has been generated by a couple of Helmoltz coils . The coils were fed by a signal generated by a wave generator Tektronix mod. AFG310 in order to achieve a magnetic field amplitude inside equal to the maximum value of the geomagnetic field measured exactly on the plate containing the samples just before the experiment start. The maximum amplitude of geomagnetic field in the laboratory ranged between 28 and 31 $\mu T$ in different experiments. These values were measured by using a F.W. Bell Sypris 5180 Gauss/Tesla meter, and the frequency was continuously monitored by an oscilloscope.

The experimental layout allows samples to be exposed to alternating fields having both a vertical and horizontal orientation rotating the plane of the coils. Since at our latitude the geomagnetic field forms a small angle (about 15°) with the vertical, we named (almost) "parallel exposure" the exposure of samples to a field generated by horizontal plane coils, like in fig.1, and (almost) "perpendicular exposure" the exposure to field generated by 90 ° tilted plane coils.

## 3. Results

Since Cayman Chemical Superoxide Dismutase Assay kit utilizes Tetrazolium salt for detection of superoxide radicals generated by XO and Hypoxhanthine, the response of both the enzymes to the e.m. exposure was tested.

In the first series of experiments, the effect of exposure on the rate of free radical generation exerted by XO was examined. XO enzyme has a $Fe_2/S_2$ (inorganic) cluster as metallic site. The exact nature of the ligands of such a structure is unknown. The Larmor frequency for this structure is $f_L = 0.54 B_0 (Hz/\mu T)$ being 0.54 the normalized *q/m* ratio for $Fe^{2+}$ ion and $B_0$ the environmental magnetic field. Figure 1 show the tetrazolium salt production as evaluated by the absorbance data: the free radical generation rate is increased in comparison with the control (the plateau value is also increased) in case of horizontal exposure, almost perpendicular to the static field.

Differences between the rates show as kinetics has a boost (in absolute value) just at the beginning of the exposure, and that the effect is markedly higher for horizontal exposure. Being the rate of the reaction linked to the concentration of the reactants, rates vanish approximately with the incubation time, as suggested by the rate equation.

In the second experimental setup it was observed that the optical density values were consistently lower than those observed in the first experiment at identical incubation time, thus indicating that SOD was able to rapidly remove a substantial percentage of generated free radicals (see Figure 2).

Since the normalized *e/m* ratio for both $Cu^{2+}$ and $Zn^{2+}$ is almost equal (*e/m* ($Cu^{2+}$) = 0.48 Hz/μT and *e/m*($Zn^{2+}$) = 0.47 Hz/μT) a frequency equal to 0.475×$B_0$ was used. Exposure to this frequency was able to modify consistently the values observed in the control run as can be seen in Figure 3. The effect being also amplified in case of an alternating field applied to horizontal configuration, i.e. almost perpendicular to static magnetic field (geo-magnetic). Since the SOD preparations we used is a Cu/Zn dependant enzyme, we ascribe the differences in the observed values to a specific action of the exposure, namely a reduced enzymatic activity of SOD.

A further check was performed in order to verify the hypothesis of a resonance effect. XO/SOD assay was exposed to different frequencies in order to compare, in the realm of the same batch, the effect of frequency. In this experiment, all the exposure were to perpendicular fields in order to maximize the effect. The sample was exposed to three different frequencies: a) the Larmor frequency; b) a lower frequency, i.e. 5 Hz ; c) a higher frequency, i.e. 325 Hz .

According to the Edmonds model, the existence of a resonant effect peaked around $f_L$ was verified. In fact, we had: Abs. rate at 15 Hz > Abs. rate at 325 Hz ~ Abs.rate at 5 Hz (see Figure 4). The observed effect (namely a decrease in the activity) was always observed strongly depending from the reciprocal orientation between static and alternating field, according to the Larmor precession theory and the NMR theory.

## 4. Discussion

The presence of a resonant frequency based on the mechanism of coupling between ELF e.m. fields and living systems has been verified. It was found that, whenever exposed to a combination of static plus alternating weak magnetic field the activity of SOD1 and XO enzymes changes when the frequency of the alternating field matches the Larmor frequency of the metallic ion of the enzyme active site. The reciprocal orientation of static and alternating magnetic field affects the strength of the results as foreseen by the Larmor theory.

The imbalance of the exposed vs. non exposed curve matches what can be expected from the rate equation: $Rate = K[A]^a[B]^b$ where A and B are the reactants and *a* and *b,* respectively, their quantities. The data obtained show that the kinetic of the reaction is obviously proportional to the reactants, provided that temperature stays constant. It is known from collision theory that molecules have to bump into each other, both with enough energy and in the right orientation, for a chemical reaction to occur. The rate constant keep into consideration "how often the molecules collide", the "fraction of molecules with enough kinetic energy to break chemical bonds", i.e. the temperature of the reaction, and the "fraction of molecules correctly oriented". Since temperature was kept constant during the experiments and the intensity of alternating field was to the highest degree "non-thermal" it can be concluded that the effect of the exposure to e.m. fields is related either to the recruitment mechanism [15] or the perturbation of an intermediate process of the reaction which requires low energy.

This concept turns into the necessity of evaluating the kinetic profile of the reaction and not only the final amount of reaction products in order to pick up the effect of the e.m. field. On the other hand, the influence of e.m. fields on enzyme activity has been widely studied with contradictory results. As an example, some author [16] reported that no biologically significant effect of weak combined ac-dc magnetic field on the binding of calcium to calmodulin has been observed on the plateau of the reaction, although other authors [17] previously reported noteworthy effects. In a different experimental setting, Salamino and coworkers [18] reported that weak magnetic fields strongly decreases the calpain catalytic activity, and linked this effect to the modified availability of $Ca^{2+}$ due to magnetic field.

The data reported here show that exposure to electromagnetic fields of well defined wavelength and geometric configuration changes consistently the rate of free radical generation and/or removal exerted by enzymes of great biological relevance. In general terms, the results indicate that it is possible to increase free radicals generation with electromagnetic fields by acting on both generation and removal of the reactive species. These observations may have important implications not only from a speculative point of view, but for applicative reasons as well. As a matter of fact variations in enzymatic activities of magnitudes similar to those observed in our experiments have been reported in various conditions related to human pathological states and/or after dietary or therapeutic interventions. In particular, estimation of SOD enzymatic activity in vivo is considered a marker of absorption/efficacy of interventions designed to increase antioxidants intake and a sign of increased pro-oxidant burden in pathological states (19-22)

In conclusion, the key points of the observations are the confirmation of the resonating effect at $f_L$ of very weak fields, apparently unable to produce any effect of significant importance in the paradigm of collision theory, and the role of the reciprocal orientation of the alternating fields (parallel or perpendicular to the static, geo magnetic field). This launches an intriguing challenge to biophysics.

# 6. FIGURE LEGEND

**Figure 1**. Determination of Xanthine Oxidase activity: the absorbance units show the variation in the production of superoxide radicals under exposure to an alternating magnetic field having a frequency of 17 Hz (Larmor frequency of the $Fe^{2+}$ ion).

**Figure 2.** Xanthine Oxidase and Xanthine Oxidase/Superoxide Dismutase activity comparison (without exposure): the absorbance units show that SOD was able to rapidly remove a substantial percentage of generated free radicals.

**Figure 3**. Determination of Xanthine Oxidase/Superoxide Dismutase activity under exposure to an alternating magnetic field of 15 Hz (Larmor frequency of $Cu^{2+}$ ion).

**Figure 4**. Determination of Xanthine Oxidase/Superoxide Dismutase activity under exposure of alternating magnetic fields having different frequencies. The direction of the alternating field was horizontal and almost perpendicular to the direction of the geomagnetic (static) field (see text for details)

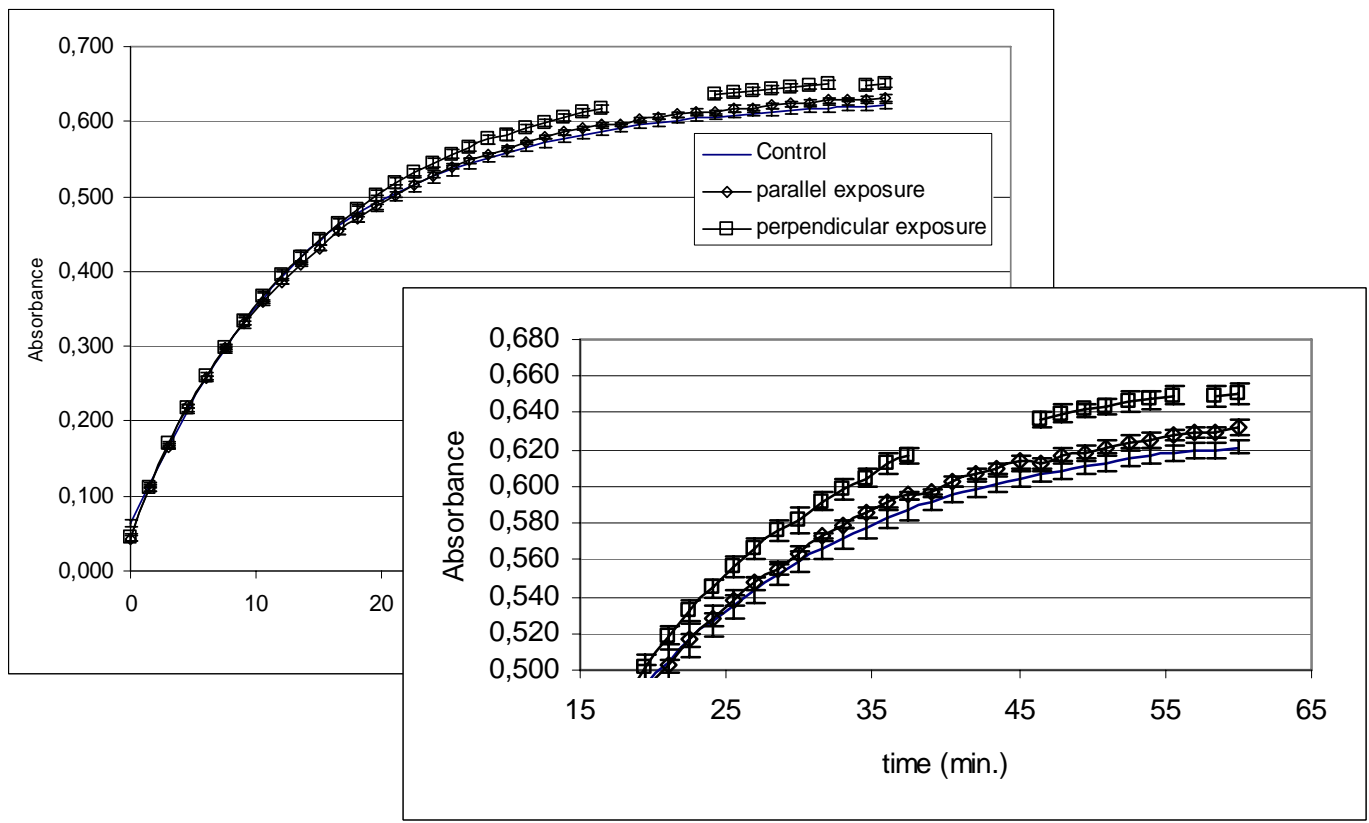

Figure 1

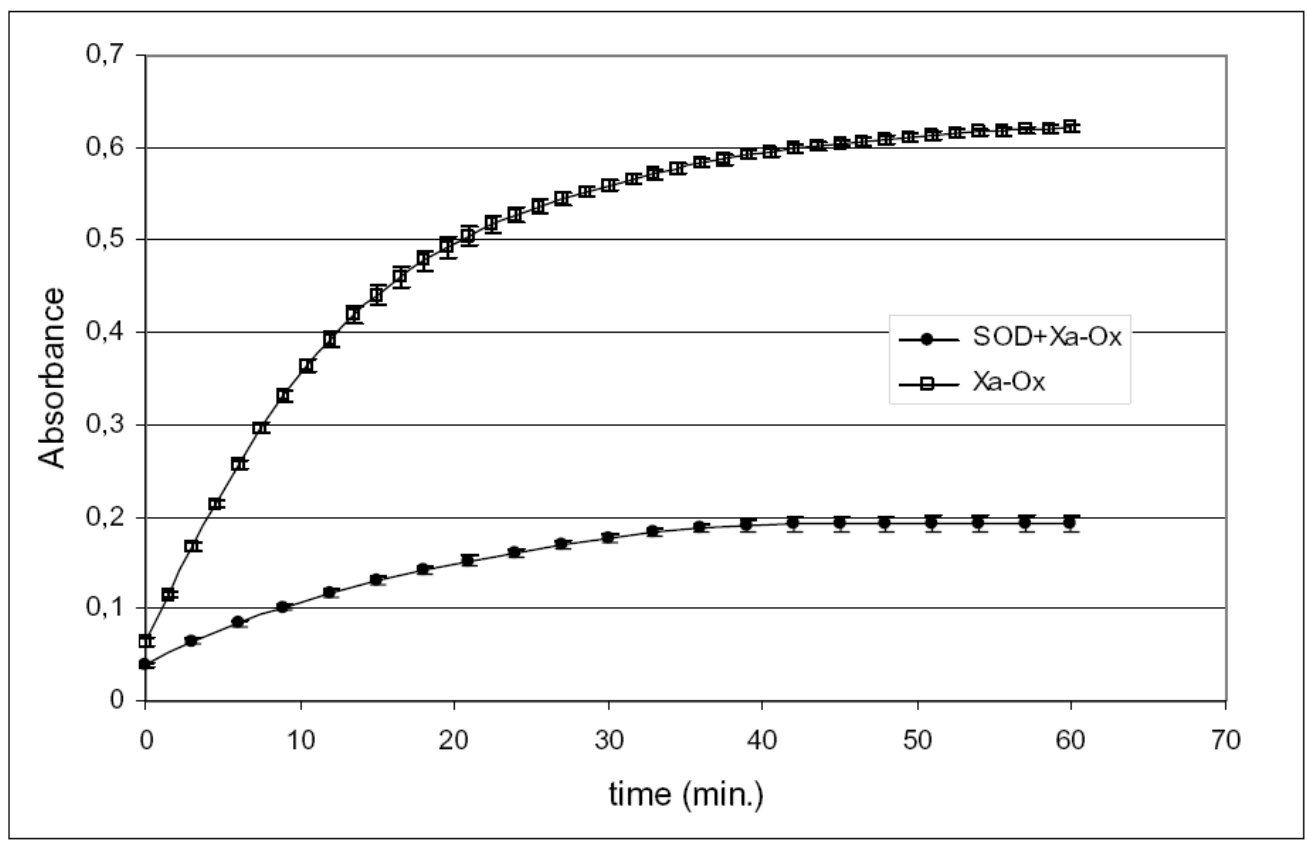

Figure 2

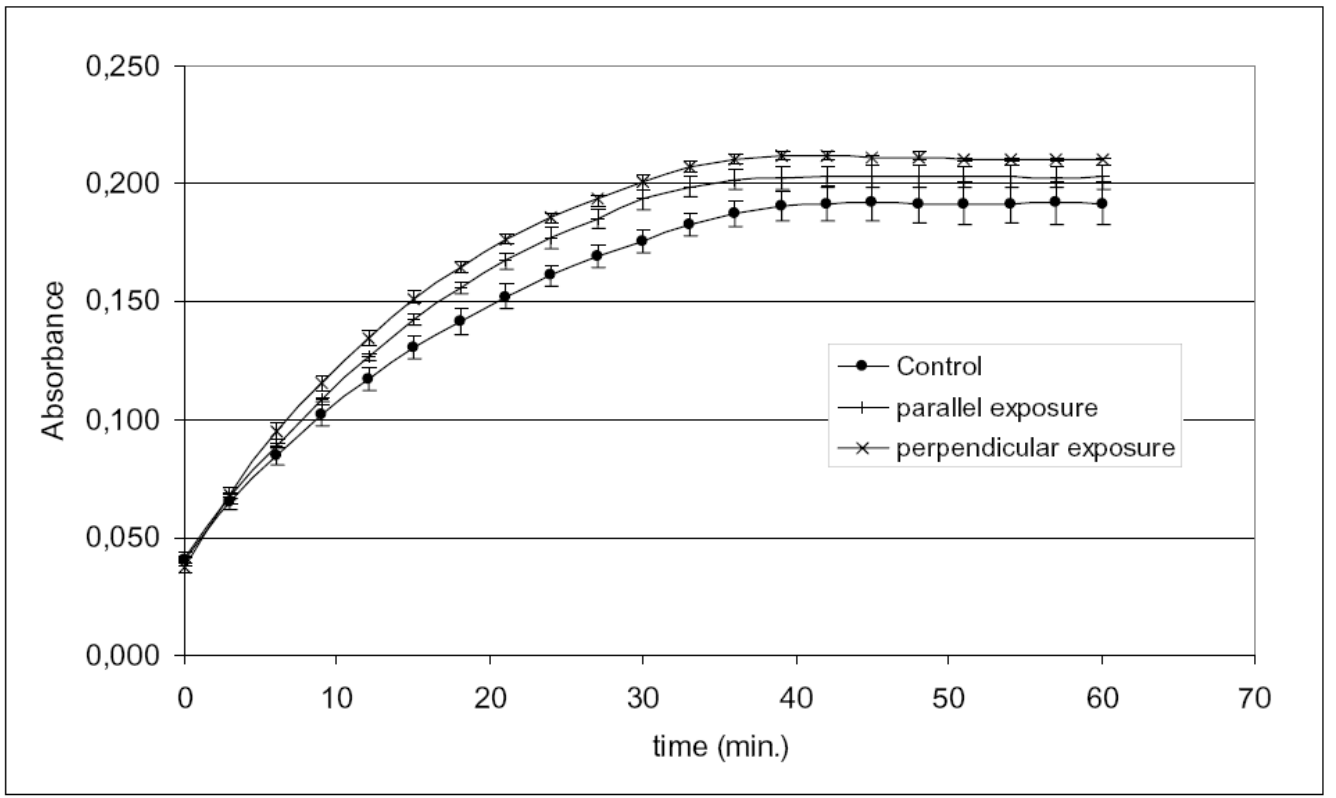

Figure 3

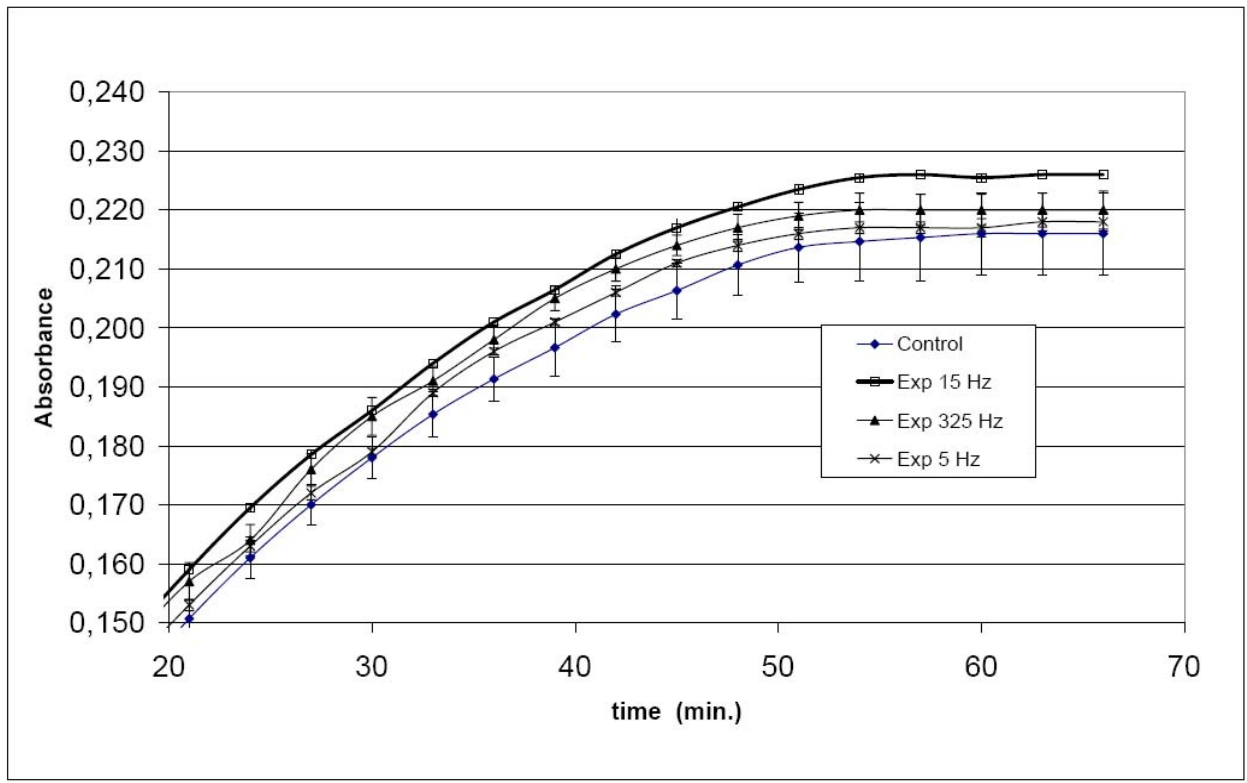

Figure 4